# On the origin of the standard genetic code as a fusion of prebiotic single-base-pair codes

A. Nesterov-Mueller*, R. Popov

Institute of Microstructure Technology, Karlsruhe Institute of Technology (KIT), Hermann-von-Helmholtz-Platz 1, 76344, Eggenstein-Leopoldshafen, Germany. *email: Alexander.Nesterov-Mueller@kit.edu

**Abstract:** The genesis of the stand genetic code is considered as a result of a fusion of two AU- and GC-codes distributed in two dominant and two recessive domains. The fusion of these codes is described with simple empirical rules. This formal approach explains the number of the proteinogenic amino acids and the codon assignment in the resulting standard genetic code. It shows how norleucine, pyrrolysine, selenocysteine and two other unknown amino acids, included into the prebiotic codes, disappeared after the fusion. The properties of these two missing amino acids were described. The ambiguous translation observed in mitochondria is explained. The internal structure of the codes allows a more detailed insights into molecular evolution in prebiotic time. In particular, the structure of the oldest single base-pair code is presented. The fusion concept reveals the appearance of the DNA machinery on the level of the single dominant AU-code. The time before the appearance of standard genetic code is divided into four epochs: pre-DNA, 2-code, pre-fusion, and after-fusion epochs. The prebiotic single-base-pair codes may help design novel peptide-based catalysts.

## 1. Introduction

The proteinogenic amino acid selenocysteine, discovered in 1974, has long been considered exceptional.[1] Indeed, it has unusual properties. Selenocysteine is encoded in a special way by the UGA codon, which is normally a stop codon and needs special translation factors.[2] But with the growth of genome sequencing capabilities, another proteinogenic amino acid pyrrolysine, a derivative of lysine, was discovered in 2002.[3,4] Pyrrolysine (Pyl) is encoded by the stop codon UAG and is an important component of enzymes in the methane metabolism in some species of prokaryotes.

These findings raised interest in puzzles known since the discovery of the standard genetic code (SGC). How many proteinogenic amino acids can exist? Is there a rational explanation for the codon multiplicity assignment? The average frequency of amino acids in proteins generally depends on the codon multiplicity, but this does not apply to all amino acids. Arginine, for example, exhibits at the same time average abundance in proteins and the largest number of codons possible for amino acids in the SGC.

It is widely accepted that the answer to these questions should be sought using the analysis of three-dimensional enzyme structures and minimization of errors during their translation.[5] Nevertheless, this approach does not find general confirmation.[6] So, the aforementioned pyrrolysine is encoded by only one codon, although this amino acid is crucial for the active site of the methyltransferase enzyme. Despite the great success in computer science on the basis of a rapidly growing database of genomes, the origin of SGC remains a mystery[7], even called the universal enigma.[8]

This work suggests that the SGC itself carries the answer to the questions stated, if individual single base-pair codes are considered as originally independent dominant and recessive codes before they merged.

## 2. Method: single base-pair codes and fusion rules

Tables 1 and 2 represent the standard genetic code but in a special form of four domains. 20 proteinogenic amino acids are distributed according to two codes consisting of two complementary pairs of bases A and U, as well as G and C. The AU-code and the GC-code can be both dominant and recessive. The dominant codes don't change after the fusion.

**Table 1:** *Dominant codes*

| | AU-code | | | GC-code | |
|---|---|---|---|---|---|
| amino acid | before fusion | SGC | amino acid | before fusion | SGC |
| **Lys** | AAA | AAA, AA**G** | **Gly** | GGG | GGG, GG**A** |
| **Asn** | AAU | AAU, AA**C** | **Gly** | GGC | GGC, GG**U** |
| **Ile** | AUU | AUU, AU**C** | **Ala** | GCC | GCC, GC**U** |
| **Phe** | UUU | UUU, UU**C** | **Ala** | GCG | GCG, GC**A** |
| **Leu** | UUA | UUA, UU**G** | **Pro** | CCC | CCC, CC**U** |
| **Tyr** | UAU | UAU, UA**C** | **Pro** | CCG | CCG, CC**A** |
| **Ile (Met)** | AUA | AUA, AU**G** | **Arg** | CGC | CGC, CG**U** |
| **Stop (Pyl)** | UAA | UAA, UA**G** | **Arg** | CGG | CGG, CG**A** |

**Table 2:** *Recessive codes*

| | AU code | | | GC code | |
|---|---|---|---|---|---|
| amino acid | before fusion | SGC | amino acid | before fusion | SGC |
| **Gln** | AAA | **C**AA, **C**AG | **Trp, Stop (Sec)** | GGG | **U**GG, **U**GA |
| **His** | AAU | **C**AU, **C**AC | **Cys** | GGC | **U**GC, **U**GU |
| **Leu (Nle)** | AUU | **C**UU, **C**UC | **Ser** | GCC | **U**CC, **U**CU |
| **Val** | UUU | **G**UU, **G**UC | **Ser** | GCG | **U**CG, **U**CA |
| **Val (Nva)** | UUA | **G**UA, **G**UG | **Thr** | CCC | **A**CC, **A**CU |
| **Asp** | UAU | **G**AU, **G**AC | **Thr** | CCG | **A**CG, **A**CA |
| **Leu (Met)** | AUA | **C**UA, **C**UQ | **Ser (X1)** | CGC | **A**GC, **A**GU |
| **Glu** | UAA | **G**AA, **G**AG | **Arg (X2)** | CGG | **A**GG, **A**GA |

The peculiarity of this construction is that the number and type of codons for each amino acid are determined according to simple empirical rules applied to the single base-pair codes. The red letters illustrate the changes according to these rules.

**Empirical rules of the single codes' fusion**
Rule 1: The codon assignment to the amino acids in the dominant codes remains after the fusion.
Rule 2: The second-position bases do not change in any code.
Rule 3: Only the third-position base changes in the dominant codes.
Rule 4: A and G, as well as U and C are exchangeable in the third position.
Rule 5: A and C, as well as U and G are exchangeable in the first position.

In brackets are amino acids that were present in the dominant and recessive codes at the early stages of the single code evolution before they merged. These cases are discussed in the next section. In the following, single base-pair codes will be called prebiotic codes.

## 3. Results and discussion
### 3.1 Dominant AU- and GC-codes
A dominant code should be considered as the older ones with respect to non-dominant ones. This is due to the fact that the dominant codes undergo less changes during code merging. As noted in the introduction, pyrrolysine is encoded with the same code UAG as the stop codon, and this code belongs to the dominant domain (Figure 2a,b). This gives an indication that pyrrolysine was present in this code in its original form.

The assumption of the appearance of stop codons as a result of the exclusion of amino acids from the code is also confirmed by other observable facts, discussed below. The reason for the exclusion of pyrrolysine from the dominant code could be the lack of reaction conditions for its conversion from lysine.

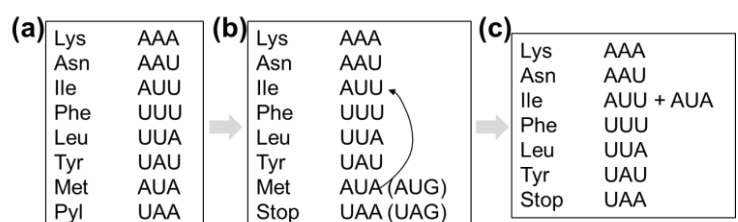

**Figure 1**: *Evolution of the dominant AU-code. (a) Initial configuration; (b) bevor DNA; (c) after DNA.*

At this stage, the dominant AU code acquired important attributes of the standard genetic code (Figure 1b): methionine (Met), which encodes the start of transcription with the AUA codon, and the UAA stop codon. These attributes boosted the self-organization of RNA and amino acids until the appearance of DNA molecules. However, the appearance of DNA greatly destabilized the system debugged during the pre-DNA era (Figure 2). The reason for this destabilization can be understood mathematically, given that the emergence of DNA-machinery has led to the additional dependence for the number of bases in accordance with the Chargaff's rules. System of equations (1) describes the number of amino acids from the dominant AU-code $Y_i$ in a protein as a function of the number of its codons $w_i$ at the DNA level:

$$\begin{cases} Y_{Lys} = w_{Lys} \\ \quad \vdots \\ Y_{Stop} = w_{Stop} \end{cases} \quad (1)$$

The system of linear equations (1), when specifying the number of amino acids, for example, through the conditions of self-organization, contains eight equations for eight independent variables $w_i$. Such a system of equations has a unique solution. In this case, the number of codons corresponds to the number of amino acids. With the introduction of the Chargaff rule, another equation appears, which follows from the equality of the number of complementary bases $N_A = N_U$ (2):

$$3W_{Lys} + W_{Asp} + W_{Met} + W_{Stop} = 3W_{Phe} + W_{Ile} + W_{Leu} + W_{Tyr} \quad (2)$$

The addition of equation (2) to the system of equations (1) leads to the dependence of the variables $w_i$. The consequence of this is the dependence of the number of one amino acid on another in a protein. This impedes the adaptation to the environmental conditions and represents a significant evolutionary disadvantage. To avoid this dependence, one of the amino acids Met was excluded from the code and amino acid Ile took its place. But since Met belonged to the dominant code, it retained the potential AUG codon, which became available only after merging prebiotic codes.

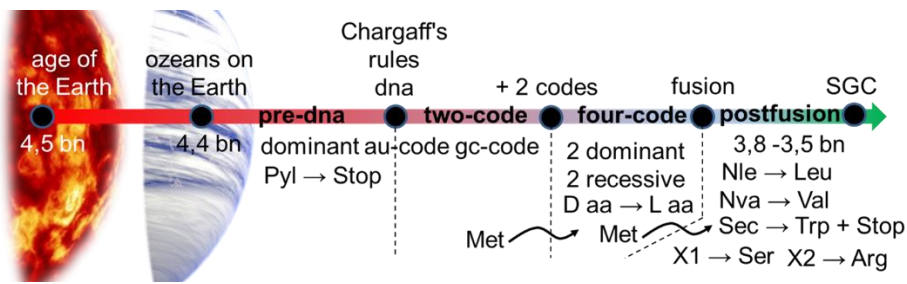

**Figure 2:** *Four prebiotic epochs in the Earth's history before the origin of the standard genetic code.*

The difficulty for estimating the moment of the DNA emergence in the dominant GC-code results from the fact that each amino acid has two codons. This excessive codon assignment does not lead to the code instability applying Chargaff's rules.

At this point, we note a remarkable fact that different amino acids were used in different codes. The reason for this could be a sequential development and a close geographical proximity of the codes, so that the number of already built-in amino acids was not enough for a long-term design of new codes.

Both dominant codes contain positively charged amino acids Lys and Arg. This contributed to the strong interaction of dominant codes with the negatively charged phosphate backbones of RNA, and presumably determined their dominance over recessive codes.

## 3.2 Recessive AU- and GC-codes

The initial recessive AU-code contained norleucine (Nle), norvaline (Nva), and methionine. Nle and Nva are abiotic amino acids. Noravalin takes the second place after valin in the synthesis yield of valine structural isomers in spark discharge experiments conducted in an environment simulating the prebiotic atmosphere on the Earth.[9] Many evidences support the hypothesis that norvaline and norleucine have been more abundant protein components during early stages of cell evolution.[10]

Met switched to the recessive code with its original AUA codon (Table 2 and Figure 2). As in the dominant AU-code, as well as in the recessive AU code, the formation of DNA led to the displacement of methionine by another representative of the leucine family - norleucine. This fact is confirmed by the mutual substitutions of Nle and Met observed in the recombinant production of proteins.[11] After the fusion of the codes, Met and Nle were replaced by Leu and Nva by Val in the SGC (Figure 2). Their mutual substitutions Leu ↔ Nle and Leu ↔ Nva were observed in the recombinant human hemoglobin produced by Escherichia coli.[12] The substitutions Leu ↔ Nva ca be explained with the sharing the same single-base-pair codon UUA by both amino acids in the prebiotic codes. Direct replacements of norvaline with valine in prokaryotes have not been reported in the literature. However, recent experiments have shown production and accumulation of Nva in a non-recombinant *E. coli* under anaerobic conditions.[13]

In the recessive GC-code, three amino acids disappeared. One of them is selenocysteine (Sel) (Table 2 and Figure 2). This disappearance has left room for two stop codons UGA and UGG. After the fusion, the UGG codon was occupied by tryptophan (Trp). Since Trp entered the SGC as a new amino acid after code fusion, it is no coincidence that Trp is the rarest amino acid in wild-type *E. coli*. Note that UGG continued to perform the function of stop codon in the mitochondria of all organisms studied till now. The reason for the disappearance of selenocysteine could be its high reactivity, which does not allow it remain in free form in cells.[14] It is known that the recessive codons of serine (AGC, AGU) and arginine (AGG, AGA) function as stop codons in the mitochondria of mammals and Drosophila.[15] This fact, as well as the heterogeneity of these codons with dominant codons of arginine, and other recessive codons of serine, suggested the existence of amino acids X1 and X2 lost from the recessive GC code after the fusion of AU- and GC-codes. Amino acid X1 had properties similar to serine. Amino acid X2 had the properties of arginine and carried a positive charge, by analogy with the other three prebiotic codes, each having one positively charged amino acid: Lys, Arg and His. Presumably, these amino acids can still be found in the proteins of extremophilic prokaryotes, characterized by a significant content of AGC, AGU, AGG or AGA codons. In particular, thermophiles and barophiles have a high AGG content (recessive Arg codon in the SGC), although dominant Arg codons are not used to increase the content of the protein stabilizing arginine.[16]

## 3.3 Principles of codes' fusion

Since the same codons are repeated in recessive and dominant codes for different amino acids, the base sequences in the codons themselves apparently had only an informative function.

Each code contained a positively charged amino acid. In the prebiotic epochs, this facilitated specific interactions between negatively charged RNA and DNA. The fusion of codes occurred after the appearance of negatively charged amino acids in a recessive AU-code. This has led to the possibility of generating more complex protein structures, as well as protein complexes from positively and negatively charged protein chains.

The question of the occurrence of stop codons is closely related to the question of the number of proteinogenic amino acids that could exist in prebiotic epochs. First, we have seen that the loss of special amino acids led to the direct occurrence of stop codons. Secondly, we made sure that the elimination of amino acids with properties similar to other amino acids in the same code can still be traced at the level of mitochondria in the form of stop codons. Thirdly, there were examples when the replacement of amino acids with structural isomers did not lead to stop codons at all. In this sense, the question remains about the amino acids in the GC-codes, which are represented by four codons instead of two (Gly, Ala, Pro, Arg, Ser and Thr). This indicates the existence of other amino acids prior to code fusion. Perhaps these were very close isomers, like the corresponding D-isomers. Since the amino acid glycine (Gly) does not have isomeric forms, it may have been represented twice in the dominant GC-code.

The maximum number of proteinogenic amino acids in prebiotic time can be estimated as 8x4 -1 (Gly) = 31 by the number of 8 codons in 4 codes. 20 of them are amino acids of the standard genetic code. The remaining 11 were distributed over four known amino acids (pyrrolysine, selenocysteine, norvaline, norleucine), two not yet known amino acids X1 and X2 predicted in this work, as well as 5 amino acids from the GC-codes.

## 4. Conclusion

The proposed concept of fusion of prebiotic codes predicts the number of biogenic amino acids and their codons in the prebiotic epochs and in the standard genetic code. The fusion concept is confirmed by the substitutions of proteinogenic amino acids from SGC observed in prokaryotes and mitochondria.

The existence of the prebiotic dominant AU-code equipped with the start and stop codons may indicate the possibility of the existence of simpler life forms beyond the SGC.

The codes' fusion concept reveals the principles of evolution to design natural enzymes, the effectiveness of which has not yet been achieved by artificial protein-based catalysts.

**Acknowledgements**. This work was supported by the DFG (no. AOBJ655892).

**Author contributions**. A.N-M. conceived the fusion of four codes to explain the standard genetic codes. A.N-M. and R.P. wrote the manuscript and sketched the four prebiotic epochs.

**Competing interests**. The authors declare no competing interests.